\documentclass[a4paper,twocolumn,11pt,accepted=2025-08-26]{quantumarticle}
\pdfoutput=1
\usepackage[utf8]{inputenc}
\usepackage[english]{babel}
\usepackage[T1]{fontenc}
\usepackage{amsmath}
\usepackage{hyperref}

\usepackage[numbers,sort&compress]{natbib}

\usepackage{graphicx}% Include figure files
\usepackage{dcolumn}% Align table columns on decimal point
\usepackage{bm}% bold math

\usepackage{cancel}
\usepackage{hyperref}
\usepackage{subcaption}
\usepackage{comment}
\usepackage{mathtools}
\usepackage{empheq}
\usepackage{bbm}
\usepackage{mathrsfs}
\usepackage{amsfonts}
\usepackage{color} 
\usepackage{pbox}
\usepackage{xcolor} 
\usepackage{braket}
\usepackage{caption}
\usepackage{enumerate}
\usepackage{physics}
\newcommand{\rcite}[1]{Ref.\;\cite{#1}}
\newcommand{\psiF}[2]{\psi_{#1}^{#2}}
\newcommand{\psiDag}[2]{\psi_{#1}^{#2\,\dagger}}
\hypersetup{
    colorlinks=true,
    linkcolor=blue,
    filecolor=magenta,      
    urlcolor=cyan,
}
\usepackage[OT2,T1]{fontenc}
\DeclareSymbolFont{cyrletters}{OT2}{wncyr}{m}{n}
\DeclareMathSymbol{\Sha}{\mathalpha}{cyrletters}{"58}

\newcommand{\be}{\begin{equation}}
\newcommand{\ee}{\end{equation}}
\newcommand{\ben}{\begin{equation*}}
\newcommand{\een}{\end{equation*}}

\newcommand{\Zb}{\mathbb{Z}}
\newcommand{\Hc}{\mathcal{H}}
\newcommand{\Oc}{\mathcal{O}}
\newcommand{\PsiB}[1]{\boldsymbol{\Psi}_{#1}}
\newcommand{\Exp}[1]{\text{Exp}\bigl[#1\bigr]}

\newtheorem{theorem}{Theorem}    

\begin{document}

\title{The Dirac Vacuum in Discrete Spacetime}

\author{Chaitanya Gupta}
 \orcid{0009-0001-9287-616X}
 \email{chai.gupta@bristol.ac.uk}
\author{Anthony J. Short}
 \email{Tony.Short@bristol.ac.uk}
\affiliation{
 H.H. Wills Physics Laboratory, University of Bristol, Tyndall Avenue, Bristol BS8 1TL, U.K}
\date{1 September 2025}

\newcommand{\add}[1]{\textcolor{teal}{\ul{#1}}}
\newcommand{\remove}[1]{\textcolor{red}{\st{#1}}}
\newcommand{\addmath}[1]{\textcolor{teal}{#1}}
\newcommand{\removemath}[1]{\textcolor{red}{\cancel{#1}}}

\graphicspath{ {./images/} }

\begin{abstract}
We consider introducing the Dirac sea in a quantum cellular automata model of fermions in discrete spacetime which approximates the Dirac equation in the continuum limit. However, if we attempt to fill up the `negative' energy states, we run into a problem. A new boundary is created between positive and negative energy states, at which pair creation seems energetically favourable.  This happens because of the modular nature of energy in discrete time models. We then suggest a possible remedy by amending the model, in order to pull states away from the new boundary. 
\end{abstract}

\maketitle

 As early as 1981, Richard Feynman suggested that a quantised version of cellular automata can be used to simulate physics \cite{feynman2018simulating}. Discrete spacetime models such as quantum walks (QW) and quantum cellular automata (QCA) have been shown to simulate some quantum field theories in the continuum limit \cite{farrelly2020review}. The evolution of the underlying quantum system in both the models is unitary and local. This raises the interesting question of whether spacetime itself may be fundamentally discrete rather than continuous. Even if this is not the case, such models may also prove useful for simulating quantum field theories on a quantum computer. This is because one can implement a QCA as a quantum circuit consisting of local unitaries \cite{schumacher2004reversible}.

While many QCA models that simulate quantum field theories have been developed\cite{destri1987light,d2017quantum,bisio2013Scattering,brun2025quantum}, interesting questions remain regarding how to introduce antiparticles in these models. The simplest models would be the Dirac QW and the Dirac QCA that simulate free Dirac particles with the former being the one-particle sector of the latter \cite{Bialynicki-Birula,succi1993lattice,TonyWalk,LeonardWalk,d2014derivation,arrighi2014dirac}. A natural way to introduce the antiparticles in these models would be to have some sort of a Dirac sea  \cite{eon2023relativistic,arrighi2020quantum,mlodinow2020quantum}. While the notion of the Dirac sea in continuum QFT is considered antiquated, it still exists in the background of the theory in the form of a simple mathematical transformation that associates the lack of a negative energy particle with an antiparticle.

 However, in this paper we argue that introducing a Dirac sea in the Dirac QCA comes with a serious problem. In particular, the modular nature of energy in discrete time leads to a new boundary between positive and negative energy states, at which pair creation could occur whilst releasing positive energy. This seems like it would cause the vacuum state to be unstable in an interacting model. To address this issue, we propose an alternative model of a Dirac QCA which pulls the energy eigenstates away from the new boundary. 

We begin by presenting the Dirac QW in section \ref{section:OgQW}. We then define a notion of energy for the QW. This allows us to introduce the Dirac sea in section \ref{sec:thediracsea} and present the problem we face in introducing the Dirac vacuum for modular energy. Then, we introduce an alternative model in section \ref{section:ModDirac}, as a possible solution to the problem. Finally, in section \ref{sec:disc}, we discuss possible future directions. 

\section{A quantum walk simulating the Dirac equation}\label{section:OgQW}

We introduce a quantum walk that has been shown to mimic the Dirac equation \cite{TonyWalk}. Consider a quantum particle on a 1-D lattice with a two-dimensional internal degree of freedom. The state of the particle is in the Hilbert space $\Hc=\Hc_{\Zb^2}\otimes\Hc_{\text{S}}$ where $\Hc_{\Zb} = \text{span}\{\ket{n}|n\in \Zb\}$ and $\Hc_{\text{S}}\cong\mathbb{C}^2$. Let $\Hc_{\text{S}}$ be spanned by the orthonormal states $\ket{r}$ and $\ket{l}$. Let the lattice spacing be $\delta x$. 

We can define the momentum eigenstates and the operator as \footnote{We have chosen the conventions $\braket{n}{m}=\delta_{n,m}$ and $\braket{p}{q}=\delta(p-q)$.}
\be
\!\ket{p} = \sqrt{\frac{\delta x}{2\pi}}\sum_n e^{i p n\, \delta x}\ket{n} \,\text{and}\,
P \!=\! \int_{-\pi/\delta x}^{\pi/\delta x} \!\!dp \, p\ketbra{p}\!,
\ee
where here and throughout this paper, we take $\hbar=1$.

We have chosen to restrict momentum from $-\pi/\delta x$ to $\pi/\delta x$ where $\pm\pi/\delta x$ are identified with each other. In a single time-step $\delta t$, the evolution of the particle is given by the unitary,
\begin{eqnarray}
    U_{\text{Dirac}}^{\text{QW}} &&= \Exp{-imc^2\sigma_x\delta t}
    \nonumber\\
    &&\quad\quad\times
    (\ket{r}\bra{r}e^{-i P\,\delta x}+\ket{l}\bra{l}e^{i P\,\delta x}) \nonumber\\ &&= \Exp{-imc^2\sigma_x\delta t}\Exp{-i P\sigma_z\delta x},
    \label{eq:DiracQW}
\end{eqnarray}

where we have defined $\sigma_z \equiv \ket{r}\bra{r}-\ket{l}\bra{l}$ and $\sigma_x\equiv \ket{r}\bra{l}+\ket{l}\bra{r}$. (We also define $\sigma_y \equiv -i\ket{r}\bra{l}+i\ket{l}\bra{r}$ which we shall use later.)

Note that $e^{-i P\delta x}$ is a shift operator, such that $e^{-i P\delta x}\ket{n}=\ket{n+1}$, hence this unitary is local, in the sense that the particle can only move a finite distance in each timestep, and hence represents a quantum walk \cite{farrelly2020review}. 

In order to connect to the standard continuous time picture, it is helpful to define a notion of energy for this system. As we are not dealing directly with Hamiltonians, we instead define the energy in terms of the eigenvalues of the unitary.  Consider the state $\ket{\psi_p}=\ket{p}\ket{s}$ where $\ket{s}$ is any state in $\Hc_{\text{S}}$. Then, consider
\be
U_{\text{Dirac}}^{\text{QW}}\ket{\psi_p} = U_{\text{Dirac}}^{\text{QW}}\ket{p}\ket{s} = \ket{p}U(p)\ket{s},
\label{eq:evolnormalboundary}
\ee
where $U(p)= \Exp{-imc^2\sigma_x\delta t}\Exp{-i p\sigma_z\delta x}$. Let $\ket{s^\pm_p}$ be eigenvectors of $U(p)$. Then, we define $E^{\pm}_p$ via
\be
    U(p)\,\ket{s^\pm_p} = e^{-iE^{\pm}_p\delta t}\ket{s^\pm_p}.
    \label{eq:eigenvec}
\ee

\begin{figure}
    \centering
    \includegraphics[width=\linewidth]{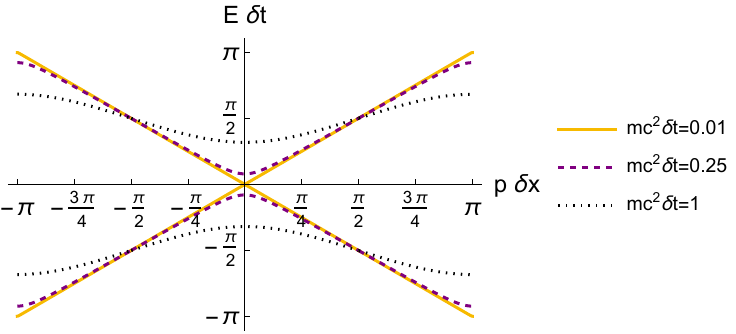}
    \caption{The dispersion relation for the 1+1-D Dirac walk. While the energy is modular $2\pi/\delta t$, we have restricted it to the range $-\pi/\delta t$ to $\pi/\delta t$. For $mc^2\delta t\ll1$, the quantum walks simulates a Dirac particle in the regime when momentum is very small (near $0$) and also in the regime when momentum is very high (near $\pi/\delta x$).}
    \label{fig:OgDiracQW}
\end{figure}

Now, we identify $E^{\pm}_p$ as the energy of the particle. It is clear that energy is modular $2\pi/\delta t$. We choose to restrict the energy from $-\pi/\delta t$ to $\pi/\delta t$ where $\pm\pi/\delta t$ are identified with each other. We have plotted the dispersion relation in Figure \ref{fig:OgDiracQW}. It can easily be seen from the figure that for every $p$, we have two energy eigenvalues with opposite signs. Hence, we make the labelling choice: $E^{+}_{p} \geq 0$ and $E^{-}_p < 0$.

Now, consider a low-momentum state of the particle $\ket{\psi_p}$ where $p \,\delta x \ll 1$ and $\ket{s}$ is any state in $\Hc_{\text{S}}$. We also suppose that $mc^2\delta t\ll 1$. Then, consider
\begin{align}
    \!U_{\text{Dirac}}^{\text{QW}}\!\ket{\psi_p} &=\! \ket{p} U(p) \ket{s}\nonumber \\ 
    &=\! \ket{p}\!(1-imc^2\sigma_x\delta t+\Oc(\delta t^2)) \nonumber \\ &\quad\quad(1-ip\sigma_z\delta x+\Oc(\delta x^2))\!\ket{s}
    \nonumber \\ 
    &=\! \ket{p}\!(1-i(mc^2\sigma_z\delta t+p\sigma_x\delta x)\nonumber\\
    &\quad\quad+\Oc(\delta x^2,\delta t^2))\!\ket{s}
    \nonumber \\
    &=\! (1-i(mc^2\sigma_z\delta t+P\sigma_x\delta x)\!\nonumber\\
    &\quad\quad+\Oc(\delta x^2,\delta t^2))\!\ket{\psi_p}
    \nonumber \\
    &\approx \!\Exp{\!-i(mc^2\sigma_z\delta t\!+\!P\sigma_x\delta x)}\!\!\ket{\psi_p}\!,
    \label{eq:diraclowexpansion}
\end{align}
where the last line is true to first order in $\delta x$ and $\delta t$.

Now, we define the effective Hamiltonian via $U_{\text{Dirac}}^{\text{QW}} = e^{-i H^{\text{eff}}\delta t}$. Comparing with (\ref{eq:diraclowexpansion}), we get the effective Hamiltonian for low momentum $H^{\text{eff}} \approx mc^2\sigma_z+P\sigma_x\delta x/\delta t$
where we have implicitly chosen a branch cut (to ensure the eigenvalues lie within $-\pi/\delta t$ to $\pi/\delta t$). Now, we fix 
\be 
   \delta t = \delta x/c
\ee
for this QW to get
\be
    H^{\text{eff}} \approx mc^2\sigma_z+Pc\sigma_x.
    \label{eq:effHamil}
\ee
This is the single-particle Dirac Hamiltonian in 1+1-D. The eigenvalues and corresponding eigenvectors of the Dirac Hamiltonian are $(E_p,\ket{\tilde{u}_p})$ and $(-E_p,\ket{\tilde{v}_p})$ where
\begin{eqnarray}
    \ket{\tilde{u}_p} \!=\! \frac{1}{\sqrt{2E_p}}(
        \sqrt{E_p+pc}\ket{r} \! + \! \sqrt{E_p-pc}\ket{l} ), \nonumber\\
    \ket{\tilde{v}_p} \!=\! \frac{1}{\sqrt{2E_p}}(\sqrt{E_p-pc}\ket{r} \!-\! \sqrt{E_p+pc}\ket{l} )
\end{eqnarray}
and $E_p=\sqrt{p^2c^2+m^2c^4}$. Moreover, for $p\,\delta x\ll1$ and $mc^2\delta t\ll1$, we can also check from Eq.~(\ref{eq:eigenvec}) with our choice of $\delta t$ that $E^{+}_{p} \approx E_p$ and $E^{-}_{p} \approx -E_p$. Also, in the same limit, $\ket{s^{+}_p}\approx\ket{\tilde{u}_p}$ and $\ket{{s}^{-}_p}\approx\ket{\tilde{v}_p}$.
Hence, for low momentum, the quantum walk mimics Dirac particles. The low momentum limit of the 1+1-D Dirac walk has been shown to mimic a free Dirac particle more rigorously in Ref.~\cite{TonyWalk,LeonardWalk,arrighi2014dirac}.

We shall now look at the high momentum limit, which typically isn't considered. Consider the state $\ket{\psi_{\pm\pi/\delta x+p}}=\ket{\pm\pi/\delta x+p} \ket{s}$ where $p\,\delta x\ll1$. Again, we also suppose that $mc^2\delta t\ll 1$.  
\begin{eqnarray}
     U_{\text{Dirac}}^{\text{QW}}\ket{\psi_{\pm\pi/\delta x+p}} &&=
     \Exp{-imc^2\sigma_x\delta t}
     \nonumber\\
     &&\quad\times\,\Exp{-i (\pm \pi/\delta x\!+\!p)\sigma_z\delta x}
     \nonumber\\
    &&\quad\quad\quad\ket{\psi_{\pm\pi/\delta x+p}}
     \nonumber\\
     &&= (-1)\, \Exp{-imc^2\sigma_x\delta t}\nonumber\\&&\quad\times\Exp{-i p\sigma_z\delta x} \ket{\psi_{\pm\pi/\delta x+p}} 
     \nonumber\\
    &&= (-1)\! \ket{\pm\pi/\delta x+p}\! U(p)\! \ket{s}\!.
     \label{eq:otherboundary}
\end{eqnarray}

Moreover, we have
\be
    E^{+}_{\pm\pi/\delta x+p} \approx \frac{\pi}{\delta t}-E_p\,, \;\; E^{-}_{\pm\pi/\delta x+p} \approx E_p-\frac{\pi}{\delta t},
\ee
$\ket{s^{+}_{\pm\pi/\delta x+p}}\approx\ket{\tilde{v}_p}$ and $\ket{s^{-}_{\pm\pi/\delta x+p}}\approx\ket{\tilde{u}_p}$. 

Comparing Eq.~(\ref{eq:otherboundary}) with Eq.~(\ref{eq:evolnormalboundary}), we can see that this is the same evolution that we had before but with an additional phase factor. Physically, the particle therefore behaves like a standard Dirac particle. However, the eigenstates for `negative' and `positive' energy are switched. (Another work has also pointed out these extra solutions and proposed a way to reinterpret them as different flavours\cite{bakircioglu2025fermion}.)

While we have introduced the Dirac walk in 1+1-D, it can also be extended to 2+1-D and 3+1-D \cite{TonyWalk,arrighi2014dirac,LeonardWalk}. We have presented the 3+1-D walk in Appendix \ref{app:3d}.
The walk with $m=0$ has also been modified to mimic the massless 1+1-D Dirac equation in curved spacetime \cite{WalkCurved}.

\section{The Dirac Sea}\label{sec:thediracsea}

\begin{figure*}
	\centering
	\begin{subfigure}{0.32\linewidth}
        \centering
		\includegraphics[width=0.75\linewidth]{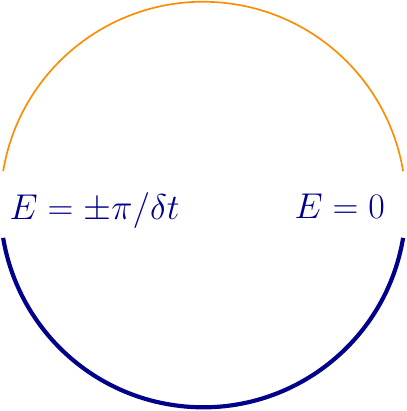}
		\caption{}
		\label{fig:EnergyCircleOG}
	\end{subfigure}
	\begin{subfigure}{0.32\linewidth}
        \centering
		\includegraphics[width=0.75\linewidth]{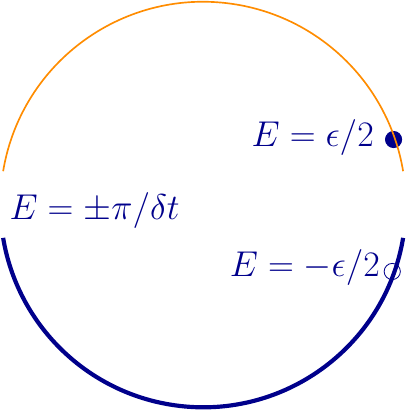}
		\caption{}
		\label{fig:PairCreation1}
	\end{subfigure}
    \begin{subfigure}{0.32\linewidth}
        \centering
		\includegraphics[width=0.75\linewidth]{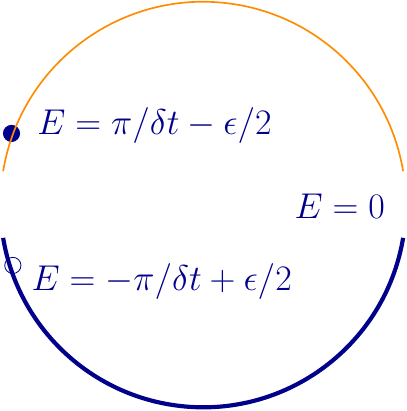}
		\caption{}
		\label{fig:PairCreation2}
	\end{subfigure}
 \caption{(a) In the Dirac QW and the Dirac QCA model, the possible energies of a particle lie on a circle as it has a modularity of $2\pi/\delta t$ with $\delta t$ being the the size of each time-step. We have also supposed that the mass is very small $mc^2\delta t \ll 1$. One way to introduce the Dirac sea is to fill up the bottom half of the energy circle (blue) while leaving the upper half of the circle empty (orange). We have two boundaries between the unfilled positive energy states and the Dirac sea \textemdash  $E=0$ and $E=\pm\pi/\delta t$. (b) In this model, a particle with energy $-\epsilon/2$ might jump to a state with energy $\epsilon/2$ creating a hole or an antiparticle at its original position. This process will require $\epsilon$ amount of energy. (c) Similarly, near the $E=\pi/\delta t$, a particle in the Dirac sea with energy $-\pi/\delta t+\epsilon/2$ might jump to a positive energy state $\pi/\delta t-\epsilon/2$ leading to pair creation. However, this process would release $\epsilon$ of energy.}
 \label{fig:energycircle}
\end{figure*}

In the continuum case, the energy spectrum of a Dirac particle runs from negative infinity to positive infinity. This means that a particle can easily `fall' to negative infinity releasing an infinite amount of energy! Dirac came up with an elegant solution \textemdash redefine the vacuum state (Dirac vacuum) as one where all negative energy states are filled with a `sea of particles' \cite{dirac1930theory}. Then, the Pauli exclusion principle stops a positive energy particle from falling to negative energies. However, now, a negative energy particle can excite to a positive energy state creating a positive energy `hole' and a positive energy particle. The hole, or the absence of a negative energy particle in the `sea of particles', can be interpreted as an antiparticle. Moreover, the Dirac vacuum defined this way is indeed the state with the lowest energy.

Mathematically, we consider $b_{p}$ and $a_{p}$ as the annihilation operators for positive and negative energy particles with some momentum $p \in \mathbb{R}$ respectively. Then, we have the bare vacuum $\ket{0}$ defined as $b_{p}\ket{0} = a_{p}\ket{0} = 0$ for all $p$. To introduce the Dirac vacuum, we introduce the annihilation operator for antiparticles $c_{p}=a^{\dag}_{-p}$. Then, we define the Dirac vacuum $\ket{\Omega}$ as $b_{p}\ket{\Omega} = c_{p}\ket{\Omega} = 0$ for all $p$. As $c_{p}=a^{\dag}_{-p}$, one can easily interpret the condition $c_{p}\ket{\Omega} = 0$ for all $p$ as having no antiparticles or having all negative energy states filled.

Can we follow the same procedure to introduce antiparticles in discrete spacetime? But what energy eigenstates do we fill up? While we have a notion of positive and negative energies in continuum QFT, it doesn't arise naturally in the Dirac QW as the dynamics are fundamentally defined by a unitary rather than a Hamiltonian. On translating the eigenvalues of the unitary into energies, we now find they lie on a circle, i.e., energy is now modular (Figure \ref{fig:EnergyCircleOG}). As in the previous section, we can choose a particular range of energies $[-\pi/ \delta t, \pi/\delta t)$  and fill up the negative energy states,  as has been suggested by Ref.~\cite{arrighi2020quantum,mlodinow2020quantum}.

Before we attempt to do so, it is imperative to have a second-quantised model. It is well known that a QW can be considered as one-particle sector of a Quantum Cellular Automata (QCA) \cite{schumacher2004reversible,vogts2009discrete,farrelly2020review}. Particularly, we can consider a fermionic QCA which is the `second quantised' version of the Dirac QW. The details of this QCA have been presented in Appendix \ref{app:QCA}. In the QCA model, one can again define $b_{p}$ and $a_{p}$ as the annihilation operators for positive energy ($0$ to $\pi/\delta t$) and negative energy ($-\pi/\delta t$ to $0$) particles with some momentum $p \in [-\pi/\delta x,\pi/\delta x)$. Similarly, we would define the annihilation operator for antiparticles $c_{p}=a^{\dag}_{-p}$. Then, we have the bare and the Dirac vacuum defined as $b_{p}\ket{0} =a_{p}ket{0} = 0$ and $b_{p}\ket{\Omega} = c_{p}\ket{\Omega} = 0$ for all $p$ respectively. Thus, the Dirac vacuum can then be obtained from the bare vacuum by filling up the states with energy lying between $-\pi/\delta t$ and $0$. 

While we are dealing the free Dirac field, we are not faced with any problems due to this definition of the Dirac vacuum in discrete spacetime. Now, suppose we want to simulate some interaction between particles (say, electromagnetic) using our Dirac QCA model. The corresponding scattering processes should conserve modular energy (We provide an argument for why modular energy should be conserved, supposing weak interactions, in Appendix \ref{sec:fermi}). Near $E=0$, we would expect to have some scattering amplitude for low momentum particle-antiparticle pair creation (Figure \ref{fig:PairCreation1}). Suppose the particle and the antiparticle created have small positive energies $\epsilon_1$ and $\epsilon_2$ respectively where $m c^2 \leq \epsilon_1, \epsilon_2 \ll 1/\delta t$. Then, the process would require the absorption of positive energy $(\epsilon_1 +\epsilon_2)$. Note that we are interested in the case when mass and energy are extremely small relative to the characteristic scales of the lattice, so that the low-momentum particles of the QCA behave like conventional continuum Dirac particles. For example, if we imagine electrons evolving on a lattice at the Planck scale then we would have $m c^2 \sim 0.5\text{MeV}$ $\ll$ $1/\delta t \sim 10^{19}\text{GeV} $. However we plot functions with larger masses in the figures so that their form is visible.  

\begin{figure*}
	\centering
	\begin{subfigure}{0.48\linewidth}
        \centering
		\includegraphics[width=\linewidth]{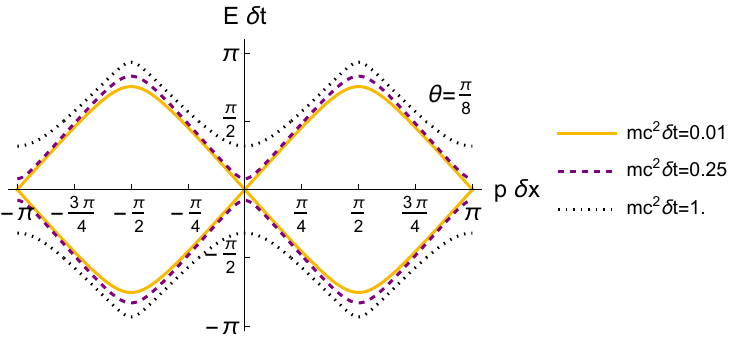}
		\caption{}
		\label{fig:ModEnergy}
	\end{subfigure}
	\begin{subfigure}{0.48\linewidth}
        \centering
		\includegraphics[width=\linewidth]{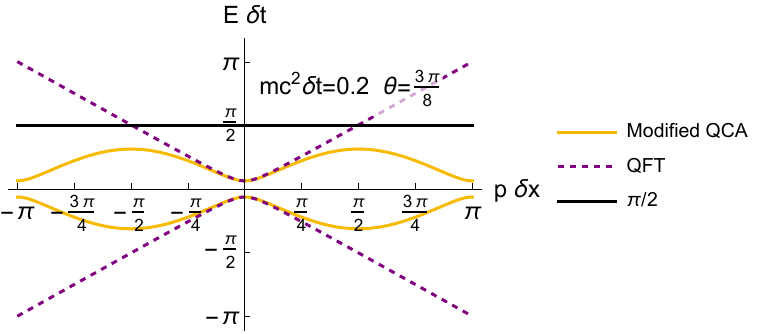}
		\caption{}
		\label{fig:ModCompare}
	\end{subfigure}
 \caption{(a) Plot of the dispersion relation for the modified quantum walk where $\theta=\pi/8$. Not that there are four momentum eigenstates for each possible energy (except at the turning points). (b) The dispersion relation for the modified quantum walk compared with the continuum QFT dispersion relation when $\theta=3\pi/8$ and $mc^2\delta t=0.2$. We can see that the energy never exceeds $\pi/2\delta t$.}
 \label{fig:ModGraph}
\end{figure*}

However, note that unlike the continuum Dirac QFT, we have two boundaries between the sea of negative energy particles and the positive energy states ($E=0$ and $E=\pm\pi/\delta t$). Hence, we would again expect to have some amplitude for pair creation pair but with the particles having high momenta (Figure \ref{fig:PairCreation2}). Suppose both the particle and the antiparticle created have positive energies $\pi/\delta t-\epsilon_1$ and $\pi/\delta t-\epsilon_2$ respectively where $m c^2 \leq \epsilon_1, \epsilon_2 \ll 1/\delta t$ again. The total energy change would then be $2 \pi/ \delta t - (\epsilon_1 + \epsilon_2)$. Due to the modular nature of energy, this corresponds to a change of $-(\epsilon_1 + \epsilon_2)<0$  if we map the energy to the desired range of $(-\pi/\delta t, \pi/\delta t]$. Hence, this process involves release of $(\epsilon_1 + \epsilon_2)$ amount of energy. This raises a potential phenomenological problem \textemdash the creation of a particle-antiparticle pair with high momenta that is energetically favourable. 

We can get further insight on what is happening by looking at the internal state of the energy eigenstates near the two boundaries. For Dirac particles, the internal state for positive energy particles is expected to be $\ket{\tilde{u}_p}$ and for the negative energy particles is expected to be $\ket{\tilde{v}_p}$. This is indeed the case for the $E=0$ boundary. However, around $E=\pm\pi/\delta t$, we have actually filled up the eigenstates of the form $\ket{\tilde{u}_p}$ \textemdash the state that we would expect for positive energy Dirac particles in continuum QFT (as we can see from the previous section). Now, in an interacting continuum theory, we do expect a Dirac particle with the internal state $\ket{\tilde{u}_p}$ to `fall down' to become a Dirac particle with internal state $\ket{\tilde{v}_p}$ as argued by Dirac himself. Thus, we should also expect this to happen at the $E=\pm\pi/\delta t$ boundary in the discrete theory if we were to introduce some interactions.

Because the eigenstates are switched and pair creation is energetically favourable, we would expect particles in the Dirac sea close to the high energy boundary to `fall up' to fill the empty positive energy states. Moreover, we might expect the amplitude for this to be the same as the magnitude for particles to `fall down' close to the original boundary without the Dirac sea because of the symmetry of the structure of internal states near both boundaries. \emph{Ironically, we have encountered a problem very similar to the original problem (in continuum QFT) which the Dirac sea was introduced to resolve.}

\section{The 1+1-D modified Dirac QW}\label{section:ModDirac}

One way to solve the problem of energetically favourable high momentum pair creation could be to pull the available energy eigenstates away from the problematic boundary between the Dirac sea and the positive energy states. In this section, we present a modified quantum walk that does so. 

\subsection{The model}

Consider the following quantum walk defined on a one-dimensional lattice with spacing $\delta x$ and time-step $\delta t$,
\begin{align}
    U_{\text{mod}}^{\text{QW}} &=\Exp{-i  m c^2 \sigma_x \delta t} \Exp{-i P\sigma_{-\theta} 
    \delta x}
    \nonumber\\
    &\quad\times
    \Exp{-i P \sigma_{\theta} \delta x }, 
    \label{eq:modwalk}
\end{align}
where $\sigma_{\theta}=R_{\theta,\hat{x}}\sigma_{z}R_{\theta,\hat{x}}^\dagger$ with $R_{\theta,\hat{n}} \equiv e^{-i\vec{\sigma}\cdot\hat{n}\theta/2}$. Note that $R_{\theta,\hat{x}}$  is simply a rotation in the internal spin and hence completely local. 

Equivalently, we have
\begin{align}
 U_{\text{mod}}^{\text{QW}} &= \Exp{-i  m c^2 \sigma_x \delta t} \nonumber \\ &\quad\times \Exp{-i P (\cos(\theta)\sigma_z+\sin(\theta)\sigma_y)\delta x} \nonumber \\ &\quad\times \Exp{-i P (\cos(\theta)\sigma_z-\sin(\theta)\sigma_y) \delta x }. 
\end{align}
It is easy to check that this QW is indeed local \textemdash an application of the unitary results into the particle moving a finite distance. This walk is similar to the twisted quantum walk defined in Ref.~\cite{jolly2023twisted}.

Again, we can define the energy of such a particle by considering the state $\ket{\psi_p}=\ket{p}\ket{s}$. We have,
\be
    U_{\text{mod}}^{\text{QW}}\ket{\psi_p} \!= U_{\text{mod}}^{\text{QW}}\ket{p}\ket{s} = \ket{p}U_{\text{mod}}(p)\ket{s},
\ee
where 
\begin{align}
    U_{\text{mod}}(p) &= \Exp{-i  m c^2 \sigma_x \delta t} \nonumber \\ &\quad\times \Exp{-i p (\cos(\theta)\sigma_z+\sin(\theta)\sigma_y)\delta x} \nonumber \\ &\quad\times \Exp{-i p (\cos(\theta)\sigma_z-\sin(\theta)\sigma_y) \delta x }.  
\end{align}
Let $\ket{\tilde{s}^\pm_p}\in\Hc_{\text{S}}$ be eigenvectors of $U_\text{mod}(p)$. Then, we define $\widetilde{E}^{\pm}_p$ as
\be
    U_{\text{mod}}(p)\,\ket{\tilde{s}^\pm_p} = e^{-i \widetilde{E}^{\pm}_p \delta t}\ket{\tilde{s}^\pm_p}.
    \label{eq:eigenvec2}
\ee
Again, we restrict the energy to the range $(-\pi/\delta t, \pi/\delta t]$ where $\pm\pi/\delta t$ are identified with each other and make the labelling choice: $\widetilde{E}^{+}_{p} \geq 0$ and $\widetilde{E}^{-}_p < 0$. Now, from Figure \ref{fig:ModEnergy}, we can see that we have four momentum  eigenstates with the same energy \textemdash twice as many as we want. This  is an example of the fermion doubling problem which is well-known in lattice field theories.  

Again, consider a low-momentum state of the particle $\ket{\psi_p}$ where $p\, \delta x \ll 1$ and $\ket{s}$ is any state in $\Hc_{\text{S}}$. We also suppose that $m c^2\delta t\ll 1$. Then, consider
\begin{align}
    U_{\text{mod}}^{\text{QW}}\ket{\psi_p} &= \ket{p} U_{\text{mod}}(p)\ket{s}\nonumber \\ 
    &= \ket{p}(1-i(mc^2\sigma_x\delta t+2p\sigma_z\cos(\theta)\,\delta x)\nonumber \\ &\quad+\;\Oc(\delta x^2,\delta t^2))\ket{s}
    \nonumber \\
    &= (1-i(mc^2\sigma_x\delta t+2P\sigma_z\cos(\theta)\,\delta x)\nonumber \\ &\quad+\;\Oc(\delta x^2,\delta t^2))\ket{\psi_p}
    \nonumber \\
    &\approx\!\Exp{-i(mc^2\sigma_x\delta t
    \nonumber\\
    &\qquad+
    2P\sigma_z\cos(\theta)\delta x)}\!\!\ket{\psi_p}\!,
    \label{eq:Moddiraclowexpansion}
\end{align}
where the last line is true to first order in $\delta x$ and $\delta t$.

With $U_{\text{mod}}^{\text{QW}} = e^{-iH^{\text{eff}}_\text{mod}\delta t}$ and fixing
\be
    \delta t = 2\cos(\theta)\,\delta x/c,
\ee
we obtain $H^{\text{eff}}_{\text{mod}}\approx Pc\,\sigma_z + mc^2 \sigma_x$
where we have implicitly chosen a branch cut. This is the Dirac Hamiltonian, and we  find $\widetilde{E}^{\pm} \approx \pm\sqrt{p^2c^2+m^2c^4}$. This is again the dispersion relation for a Dirac particle. We have presented a fermionic QCA that is the `second quantised' version of this model in Appendix \ref{app:QCA}.

\subsection{The Energy Spectrum and the Dirac Sea}

\begin{figure}
        \centering
		\includegraphics[width=0.60\linewidth]{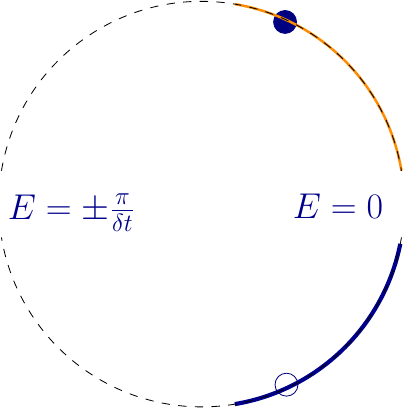}
		\caption{While energies still lie on a circle for the modified QCA, not all energy states on the circle are valid. One can again fill up all the `negative' energy states (blue) and let the `positive' energy states (orange) be unfilled. By pushing the possible energy states far back enough, we can make the gap at the $E=\pm\pi/\delta t$ larger than half of the circle. If pair creation happens at the states closest to the $E=\pm\pi/\delta t$ boundary, it will still involve absorption of `positive' energy.}
	\label{fig:PairCreationMod}
\end{figure}

We can again define the Dirac sea by filling up all the negative energy states.  Again, a negative energy particle might excite to a positive energy state leading to pair creation. However, the parameter $\theta$ allows us to pull energies away from the problematic boundary ($E=\pm\pi/\delta t$) and create a gap. The high-energy particles are no longer in the same region where we would expect them to have the dynamics of a Dirac particle. Due to the energy gap, we might expect the amplitude for high-momentum particle creation to be damped. Moreover, we have the following theorem which we prove in Appendix \ref{app:theorem}.

\begin{theorem}
    For any $0\leq mc^2\delta t<\pi/2$, there exists a $\theta$ such that $\abs{\widetilde{E}^{\pm}_p}\delta t<\pi/2$. 
    \label{thm:theta}
\end{theorem}

In other words, we can choose an appropriate value of $\theta$ such that the gap between the positive and negative energy states at the problematic boundary is larger than $\pi/\delta t$ as in Figure \ref{fig:ModCompare}. Then, pair creation never leads to release of `positive' energy. In particular in order for a particle to 'jump' from anywhere in the Dirac sea to a positive energy state would require the absorption of energy in the range $2mc^2<E<\tfrac{\pi}{\delta t}$ from some other field, rather than emitting positive energy into that other field.

Finally, it is also interesting to note that we do not get an energy gap at the problematic boundary for all possible values of $\theta$. In particular, when $2\theta = mc^2 \delta t$, the mass gap at the $E=\pm\pi/\delta t$ boundary vanishes. 

\section{Discussion}\label{sec:disc}

Although we have focussed on a 1+1-D model in the main body of the paper for simplicity, the modular nature of energy remains in higher dimensional models. The models we have considered can be naturally extended to such cases and similar issues regarding the Dirac sea occur. One important difference in higher dimensions is that even the simplest Dirac quantum walks involve fermion doubling. We have discussed the 3+1-D case in Appendix \ref{app:3d}.

A very interesting paper \rcite{eon2023relativistic} recently presented a quantum cellular model of quantum electrodynamics in 2+1-D and 3+1-D. In a section of this paper on pair creation, the authors introduce a construction of the Dirac sea in which all the antiparticle degrees of freedom are populated. This is associated with a product state in which exactly half of the fermionic modes at each site are occupied. However, we would expect that, although the Dirac sea has a simple structure in momentum and energy space, it would have a highly entangled structure in position and not be represented by a separable state (as shown by the vacuum correlation functions in continuum QFT \cite{peskin2018introduction}). Another limitation of this construction is that the unitary in the QCA for the mass term seems to be causing pair production (even when we set the electromagnetic interaction off by setting the coupling constant to zero in the model). It would be very interesting to consider alternative constructions of the Dirac sea in this QCA model, which are closer to those we consider here, to simulate QED. 

We argued earlier that we would expect pair creation at the high energy boundary of the Dirac sea in a discrete time model, based on energy conservation arguments, in a similar spirit to Dirac's original argument \cite{dirac1930theory}. Additionally, we also show that we should have modular energy conservation if we were to deal with an interacting QCA perturbatively (Appendix \ref{sec:fermi}). In this paper, we have provided arguments on why the aforementioned problem is likely to exist; it would be very interesting to calculate such effects from an explicit interacting model, which could be developed either by adding an additional unitary interaction to the quantum cellular automata, or by adding additional physically motivated fields. One approach would be to look at free photon QCA models\cite{bisio2016quantum,d2017quantum,brun2025quantum}. However, in such models, we do not yet have a way to couple fermions with photons using a local unitary that gives the correct interaction term in the continuum limit. Alternatively, we can look at QCAs simulating Quantum Electrodynamics in 1+1-D and 3+1-D \cite{arrighi2020quantum,sellapillay2022discrete,eon2023relativistic}. While these models do simulate interactions between a gauge field and Dirac particles, we do not have an idea on how to write the free vacuum state or photon states in such models. When introducing the Dirac sea in these models, we would also have to be careful to distinguish the free Dirac vacuum from the interacting vacuum \textemdash a distinction that also exists for continuum QFT and is not unique to QCA models\footnote{Suppose we have a Hamiltonian $H=H_0+V$ where $H_0$ determines the free evolution and $V$ the interaction. Then, the free vacuum (for instance, the Dirac vacuum) is the lowest energy eigenstate of $H_0$ and the true interacting vacuum is the lowest energy eigenstate of $H$.}.  

 In order for the quantum cellular automata models to simulate the physics we observe, it is important that a relatively stable total energy can be defined. Due to the modular nature of energy, this means that interaction terms which would generate $2 \pi/ \delta t $ of energy (with respect to the free particle model) should be suppressed. This is a key problem for discrete time approaches to  resolve. 

While our new model offers a possible approach to eliminating pair creation at the high energy boundary of the Dirac sea, it comes at the cost of fermion doubling. This is something that has been dealt with in lattice field theories (LFTs) \cite{wilson1974confinement,kogut1975hamiltonian,susskind1976strong,montvay1994LFT,rothe2012lattice}. It would be interesting to investigate whether similar approaches would work in our model, or perhaps if any doublers could be interpreted as different particle flavours.   

Alternatively, perhaps a better approach that does not lead to fermion doubling might be possible. Fundamentally, the problem with finding the ground state or the Dirac vacuum in discrete time models lies with the modular nature of energy in these models. In continuum QFT, a true vacuum state is defined with respect to a Hamiltonian as being the eigenstate with the lowest energy. However, due to the modular nature of energy, it is somewhat arbitrary to have a notion of a lowest energy state in discrete time models. Should we abandon the idea of introducing a Dirac sea and attempt to introduce antiparticles in another way? 

In conclusion, our exploration of the Dirac sea in discrete spacetime raises a number of potential issues. Although we have proposed a possible resolution, there remain many interesting avenues for further investigation. 

\section{Acknowledgements}
We thank Charlie Shakeshaft and Peter Walters for their insights on the points in momentum space around which the original 3+1-D Dirac walk behaves like two Weyl particles.

\appendix

\section{Modular Energy Conservation: Discrete Fermi's Golden Rule}
\label{sec:fermi}

We wish to show that, in a perturbative treatment of an interacting QCA, modular energy, the way we have defined it, is conserved. We shall derive a discrete version of Fermi's golden rule in this section. Consider a QCA described by $U=TV$ where $T$ is the free evolution, and $V$ dictates the interactions. Let $V= \Exp{-i \delta t \alpha H}$ where $\alpha$ is the coupling constant and $H$ is the effective interaction Hamiltonian.

Suppose $\ket{\psi_f}$ and $\ket{\psi_i}$ are two distinct eigenstates of $T$ with eigenvalues $e^{-iE_i \delta t}$ and $e^{-i E_f \delta t}$.

Assuming that the interaction is weak, for time $\tau = N\delta t$, we have

\begin{align}
    \bra{\psi_f}(TV)^N\!\!\ket{\psi_i} &=  \bra{\psi_f}T^N\ket{\psi_i}
    \nonumber\\
    &\quad- i \delta t\, \alpha\! \sum_{n=0}^{N-1} \bra{\psi_f}\!T^{N-n}\!HT^{n}\!\ket{\psi_i} 
    \nonumber\\
    &\quad+\Oc(\alpha^2).
    \label{eq:firstorder}
\end{align}

Now, we wish to calculate the transition rate defined as
\be
 \Gamma_{i \to f} = \lim_{N\to \infty} \frac{1}{N \delta t} \abs{\bra{\psi_f}(TV)^N\!\!\ket{\psi_i}}^2.
\ee

Considering only the first order term, we can write the transition rate as
\begin{align}
    \Gamma_{i\to f}^{(1)}&\!=
    \!\lim_{N\to\infty}\!\frac{1}{N\delta t} \abs{- i \delta t\, \alpha\! \sum_{n=0}^{N-1} \bra{\psi_f}\!T^{N-n}\!HT^{n}\!\ket{\psi_i}}^2 
 \nonumber\\
 &=  \lim_{N\to\infty}\frac{1}{N \delta t} \alpha^2 \delta t^2 \abs{\sum_{n=0}^{N-1} e^{i n (E_f - E_i) \delta t}}^2 
 \nonumber\\
 &\quad\times
 \abs{\bra{\psi_f}H\ket{\psi_i}}^2.
    \label{eq:transitionInt}
\end{align}

Now, we shall use the following identity;
\be
    \lim_{N\to\infty} \frac{1}{N T} \abs{\sum_{n=0}^{N} e^{i 2 \pi x n / T}}^2 = \sum_{k \in \Zb} \delta (x-k T) 
\ee
where the expression on the right hand side is known as the Dirac comb. Thus, we get

\be
    \Gamma_{i\to f}^{(1)} = 2 \pi \alpha^2\,  \abs{\bra{\psi_f}H\ket{\psi_i}}^2 \sum_{k \in \Zb} \delta (E_f-E_i-\tfrac{2\pi}{\delta t}k).
\ee

Hence, we have obtained a discrete version of Fermi's golden rule which has modular energy conservation. We expect the higher order terms to also similarly give modular energy conservation.
\section{Fermionic Quantum Cellular Automata}\label{app:QCA}

We shall introduce the 1+1-D Dirac QCA in this section \cite{mauro2010quantum,d2012physics,Bialynicki-Birula}. We present the fermionic rendition of this QCA and discuss the mapping to qubits in the next section. Again, consider a 1-D lattice with spacing $\delta x$. At each lattice point $n\in \Zb$, we introduce the operators $\psiF{n}{r}$ and $\psiF{n}{l}$ defined at each lattice point ${n}\in\Zb$ such that 
\be
\{\psiF{n}{a}\,,\psiDag{m}{b}\}=\delta_{nm}\delta^{ab}, 
\ee
where $a,b\in\{r,l\}$ and ${n},{m}\in\Zb^2$. We shall write
\be
    \PsiB{n} = \begin{pmatrix}
        \psiF{n}{r} \\
         \psiF{n}{l}
    \end{pmatrix}.
\ee

To look at the momentum space, we define the Fourier transform of the two-component field as follows
\be
\PsiB{p} = \sqrt{\frac{\delta x}{2 \pi}}\sum_{n\in\Zb}\PsiB{n}e^{-ipn\,\delta x},
\ee
where we again restrict $p$ from $-\pi/\delta x$ to $\pi/\delta x$.

Consider the unitary $T$ such that
\be
    T\psiF{n}{r}T^\dag = \psiF{n+1}{r} \;\;\text{and}\;\; T\psiF{n}{l}T^\dag = \psiF{n-1}{l}.
    \label{eq:DiracQCAT}
\ee
Also, we define the unitary $W$ such that
\begin{align}
    W\psiF{n}{r}W^\dag &= \cos(mc^2\delta t)\,\psiF{n}{r}+i\sin(mc^2\delta t)\,\psiF{n}{l} \nonumber\\
    W\psiF{n}{l}W^\dag &= i\sin(mc^2\delta t)\,\psiF{n}{r}+\cos(mc^2\delta t)\,\psiF{n}{l}.
    \label{eq:DiracQCAW}
\end{align}
where $\delta t = \delta x/c$ again.
We can write the unitaries as,
\begin{align}
    W &= \Exp{-imc^2\delta t \,\! \sum_{n\in\Zb}\PsiB{n}^\dagger\,X \PsiB{n}}\nonumber\\
    T &= \Exp{-i\,\delta x\!\int_{-\pi/\delta x}^{\pi/\delta x}dp \,p\, \PsiB{p}^\dagger\, Z\PsiB{p}}.
\end{align}
where $X$ and $Z$ are Pauli-X and Pauli-Z matrices respectively.
Now, consider the following unitary that evolves the system in one time-step $\delta t$ and defines the Dirac QCA:
\begin{align}
    U^{\text{QCA}}_\text{Dirac} = WT.
\end{align}

Let us consider its single particle limit. First, we define the bare vacuum such that $\psiF{p}{a}\ket{0} = 0$ for all $a \in \{r,l\}$. Consider the state $\ket{p,a} = \psiDag{p}{a}\ket{0}$. Then, one can check how $U^{\text{QCA}}_\text{Dirac}$ acts on $\ket{p,r}$ and $\ket{p,l}$ using Eq.~(\ref{eq:DiracQCAT}) and Eq.~(\ref{eq:DiracQCAW}). Also, we identify $\ket{p,a}$ in the QCA model with $\ket{p}\ket{a}$ in the QW model. Then, comparing with Eq.~(\ref{eq:DiracQW}), one can easily see that the action of $U^{\text{QCA}}_\text{Dirac}$ on $\ket{p,a}$ is the same as the action of $U^{\text{QW}}_\text{Dirac}$ on $\ket{p}\ket{a}$. Hence, the 1+1-D Dirac QW is the single particle sector of the 1+1-D Dirac QCA.

Furthermore,  looking at the Dirac Hamiltonian we see that the Dirac QCA unitary is the Trotterisation of the Hamiltonian for free Dirac particles in 1+1-D QFT. Hence, one would expect convergence to this Hamiltonian under the continuum limit ($\delta x,\delta t\to0$). The 1+1-D Dirac QCA has been extended to higher dimensions and rigorously shown to simulate free Dirac particles \cite{farrelly2014causal,bisio2015quantum,mlodinow2020quantum,brun2020quantum,mlodinow2021fermionic}. 

We now define $a_p$ and $b_p$ that we used in Section \ref{sec:thediracsea}. By comparing the single particle sector of the QCA with the QW model, we find 
\be
\label{eq:defineOp}
\begin{pmatrix}
   \psiF{p}{r}\\[0.1cm]
   \psiF{p}{l}
\end{pmatrix}
=
\begin{pmatrix}
    \braket{r}{s^{+}_p} & \braket{r}{s^{-}_p}\\[0.1cm]
     \braket{l}{s^{+}_p} & \braket{l}{s^{-}_p}
\end{pmatrix}
\begin{pmatrix}
    a_p\\[0.1cm]
    b_p
\end{pmatrix}.
\ee

We  also define 
\begin{align}
    \boldsymbol{\sigma}^+(p) &= \sqrt{2 E_p}\begin{pmatrix}
    \braket{r}{s^+_p}\\[0.1cm]
    \braket{l}{s^+_p}
\end{pmatrix}, \nonumber\\
    \boldsymbol{\sigma}^-(-p) &= \sqrt{2 E_p}\begin{pmatrix}
    \braket{r}{s^-_p}\\[0.1cm]
    \braket{l}{s^-_p}
\end{pmatrix},
\label{eq:transform1}
\end{align}
such that 
\be
    \PsiB{p} = \frac{1}{\sqrt{2E_p}} \left( \boldsymbol{\sigma}^{+}(p)\,b_p + \boldsymbol{\sigma}^{-}(-p)\,a_p \right).
\ee
To shift from the bare vacuum to the Dirac vacuum, we must interpret the lack of a negative energy particle as an antiparticle. Hence, we make the substitution $c_p^\dag=a_{-p}$. And after taking the Fourier transform, we get the equation
\begin{align}
     \PsiB{n} = \int_{-\pi/\delta x}^{\pi/\delta x} \frac{dp}{\sqrt{2E_p}} \bigr( \boldsymbol{\sigma}^{+}(p)\,b_p e^{ipn\,\delta x} + \nonumber\\
     \boldsymbol{\sigma}^{-}(p)\,c_p^\dag e^{-ipn\,\delta x} \bigl).
    \label{eq:QCAModeExpand}
\end{align} 

Let us look at the mode expansion for 1+1-D continuum Dirac QFT. Defining 
\begin{align}
    \mathbf{u}(p) = \sqrt{2E_p}\begin{pmatrix}
    \braket{r}{\tilde{u}_p}\\[0.1cm]
    \braket{l}{\tilde{u}_p}
    \end{pmatrix}
    \;\text{and}
    \nonumber\\
   \mathbf{v}(-p) = \sqrt{2E_p}\begin{pmatrix}
    \braket{r}{\tilde{v}_p}\\[0.1cm]
    \braket{l}{\tilde{v}_p}
    \end{pmatrix}, 
\end{align}
we can write the mode expansion as
\be
    \boldsymbol{\Psi}(x) \!\!=\!\! \int_{-\infty}^{\infty}\!\! dp\frac{1}{\sqrt{2E_p}}\left( \mathbf{u}(p)\,b_p e^{ip x} + \mathbf{v}(p)\,c_p^\dag e^{-ip x} \right).
    \label{eq:modeExpandStd}
\ee

This looks very similar to the mode expansion in the QCA from Eq.~\eqref{eq:QCAModeExpand} . Moreover, 
for small $p$, we have already shown that $\boldsymbol{\sigma}^{+}(p)\approx\mathbf{u}(p)$, and $\boldsymbol{\sigma}^{-}(p)\approx\mathbf{v}(p)$, hence for small momentum the models align. For states with momentum close to $\pm\pi/\delta x$, we have
$\boldsymbol{\sigma}^{+}(\pm\pi/\delta x+p)\approx\mathbf{v}(p)$ and $\boldsymbol{\sigma}^{-}(\pm\pi/\delta x+p)\approx\mathbf{u}(p)$.

Now, we can also `second quantise' the modified quantum walk. We present the modified Dirac QCA. Consider

\begin{align}
    T_{\theta} &= \text{Exp}\bigl[ -i\,\delta x \!\int_{-\pi/\delta x}^{\pi/\delta x}\!\!\!\!dp\,\PsiB{p}^{\dagger}\,p\,(\cos(\theta)Z
    \nonumber\\
    &\qquad\qquad-\sin(\theta)Y)\PsiB{p} \bigr],
\end{align}
where Y is the Pauli-Y matrix and, as before, we have
\begin{align}
 W &= \Exp{-imc^2\delta t \sum_{n\in\Zb}\PsiB{n}^\dagger\,X\PsiB{n}}.
\end{align}

Then, we introduce the modified Dirac QCA,
\be
    U^{\text{QCA}}_{\text{mod}} = WT_{-\theta}T_{\theta}.
\ee

Again, the annihilation and creation operators for positive and negative energy particles can be similarly defined as in Eq.~\eqref{eq:defineOp}. This allows us to define the Dirac sea more formally being in the `second quantisation' picture.

So far, we have not been dealing with standard QCAs, but rather fermionic QCAs. To simulate our QCAs on a quantum computer, we must look at its circuit representation where the fermionic degrees of freedom are mapped to qubits.

\section{Quantum Circuits for 1+1-D Fermionic QCAs}\label{app:JW}

In 1+1-D, we can map fermions to qubits through a Jordan-Wigner transformation while maintaining locality \cite{farrelly2014causal,jordan1993paulische}. Suppose there are two qubits at each lattice site to represent the occupation number of right and left moving fermions.  Let the qubits be labelled by $(n,a)$ where $n$ is the lattice site and $a\in\{r,l\}$ is the chirality. Define $A_{n}^{a\dag}=\ket{1}\bra{0}_{n}^{a}$ where $a\in\{r,l\}$. Then, the Jordan-Wigner transformation can be written as 
\begin{eqnarray}
    \psi^{r\dag}_n &= A_{r}^{r\dag}Z_{n}^{l}Z_{n-1}^{r}Z_{n-1}^{l}\dots,
    \nonumber\\
    \psi^{l\dag}_n &= A_{l}^{l\dag}Z_{n-1}^{r}Z_{n-1}^{l}Z_{n-2}^{r}\dots,
    \label{eq:JW}
\end{eqnarray}
where $Z_{n}^{a}=\ket{0}\bra{0}-\ket{1}\bra{1}$ are just Pauli-Z operators acting on the respective qubits. This ensures that anti-commutativity is preserved.

\subsection{Dirac QCA}

\begin{figure}[h!]
    \centering
    \includegraphics[width=0.9\linewidth]{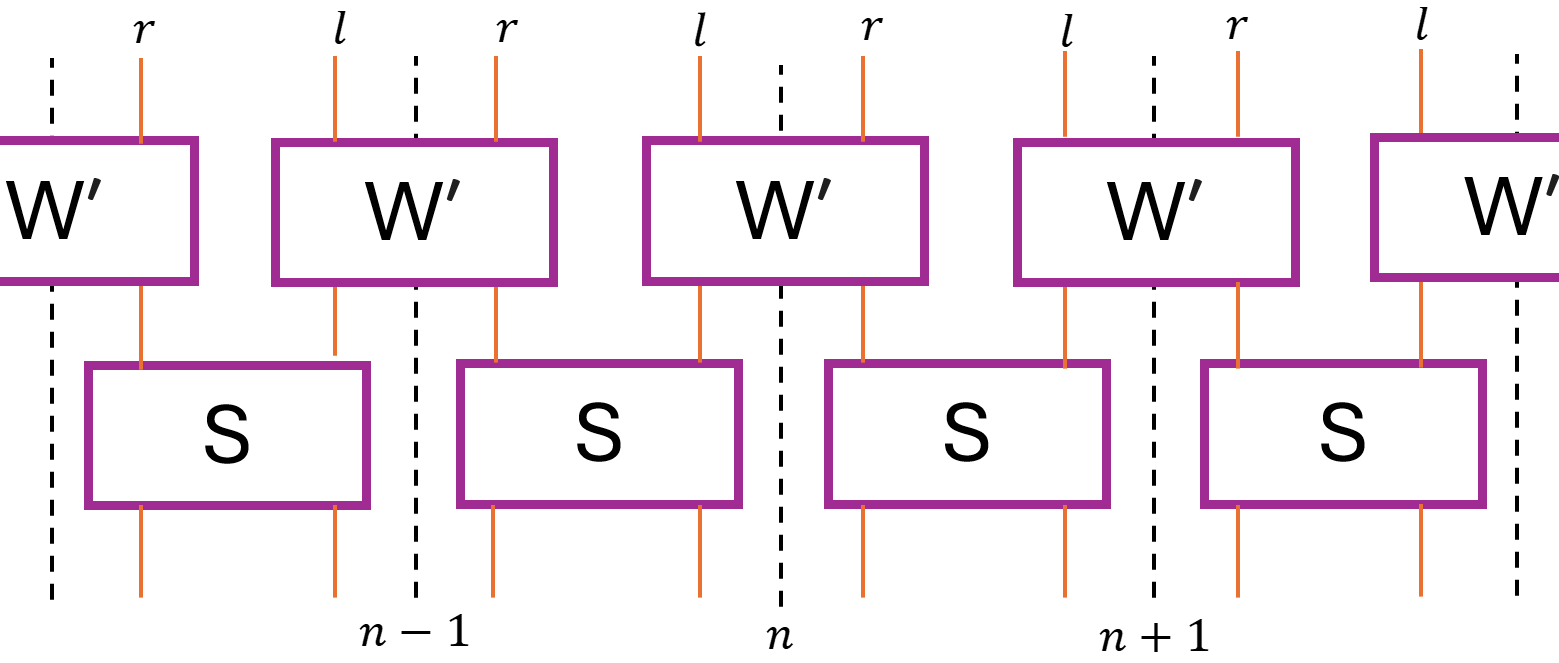}
    \caption{The Dirac QCA can be decomposed into local unitaries acting on qubits. The circuit is equivalent to $U_{\text{Dirac}}^{\text{QCA}}$ under Jordan-Wigner transformation as defined in Eq.~(\ref{eq:JW}).}
    \label{fig:OGCircuit}
\end{figure}

The circuit representation of the Dirac QCA has been developed in Ref.~\cite{farrelly2014causal,arrighi2020quantum}. We present an equivalent circuit. Suppose $s_m=\sin(mc^2\delta t)$ and $c_m=\cos(mc^2\delta t)$. 

Consider the gates $S$ and $W'$ defined as
\begin{align}
    S = \begin{pmatrix}
        1 & 0 & 0 & 0 \\
        0 & 0 & 1 & 0 \\
        0 & 1 & 0 & 0 \\
        0 & 0 & 0 & -1 
    \end{pmatrix}    \;\text{and}
    \nonumber\\
    W' = \begin{pmatrix}
        1 & 0 & 0 & 0 \\
        0 & c_m & -is_m & 0 \\
        0 & -is_m & c_m & 0 \\
        0 & 0 & 0 & -1 
    \end{pmatrix},   
    \label{eq:OgCircuit}
\end{align}
where $S$ acts on qubits $(n,r)$ and $(n+1,l)$ and is written in the computational basis $\{\ket{00},\ket{01},\ket{10},\ket{11}\}$ where $\ket{ab}_{n,n+1}=\ket{a}_{n}^{r}\ket{b}_{n+1}^{l}$, and $W'$ is also written in the computational basis $\{\ket{00},\ket{01},\ket{10},\ket{11}\}$ but $\ket{ab}_{n}=\ket{a}_{n}^{l}\ket{b}_{n}^{r}$.
(Here, $S$ can be described as fermionic swap.)

Then, the unitary $U_{\text{QCA}}$ for the Dirac QCA can be represented as shown in Figure \ref{fig:OGCircuit}. Using the transformation described in Eq.~(\ref{eq:JW}), one can easily check the circuit is indeed equivalent to $U_{\text{QCA}}$. Finally, for simulation on a quantum computer, one needs to consider a finite lattice with $N$ lattice sites and periodic boundary conditions such that site $N$ is identified with site $0$.

\subsection{Modified Dirac QCA}
\begin{figure}[h!]
    \centering
    \includegraphics[width=0.9\linewidth]{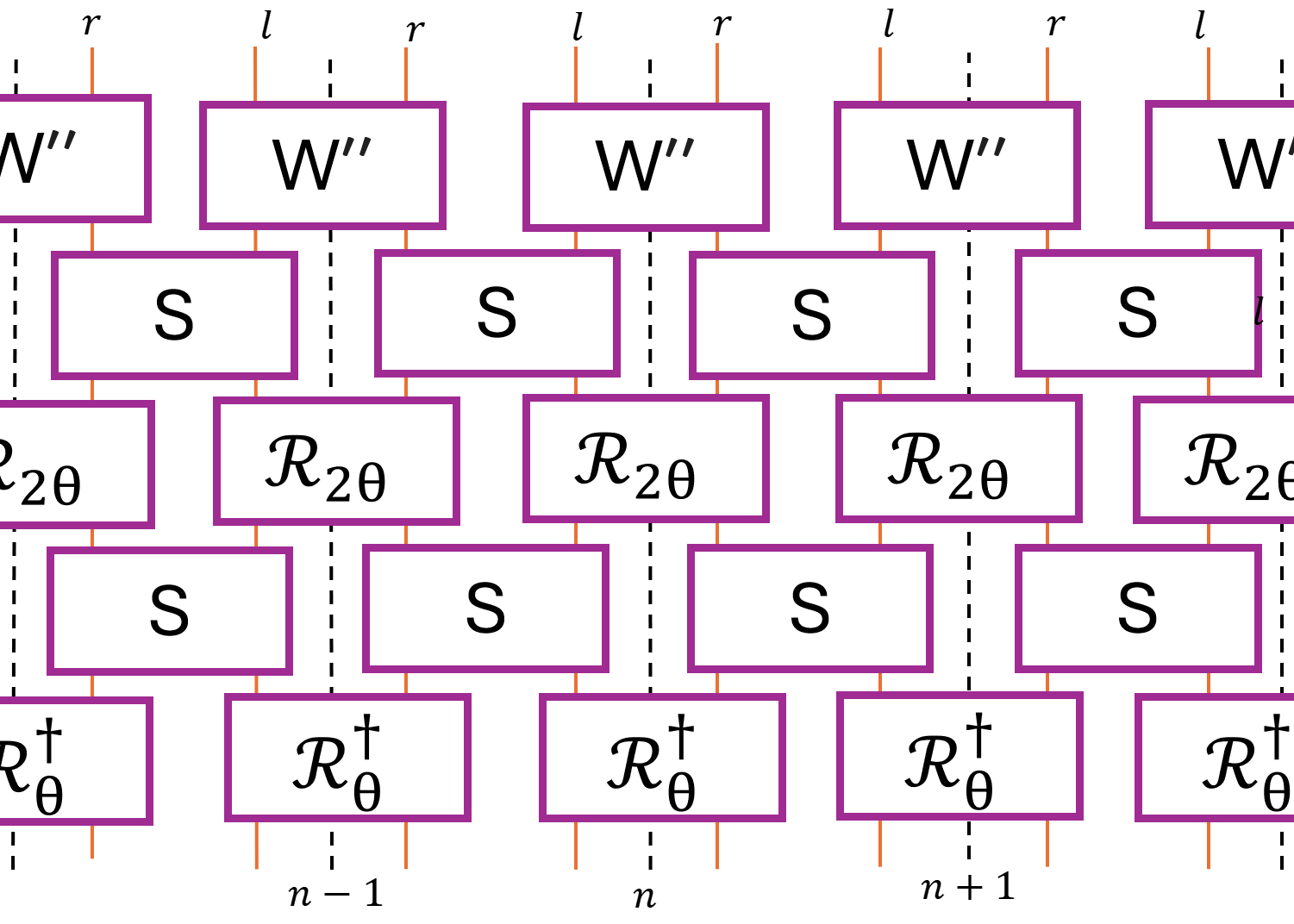}
    \caption{The circuit representation for the fundamental unitary of our modified Dirac QCA. Here, we have mapped fermionic degrees of freedom to qubits using transformations defined in Eq.~(\ref{eq:JW}).}
    \label{fig:ModCircuit}
\end{figure}

Now, we wish to have a circuit representation for the modified QCA. Consider the following gate acting on qubits $(n,r)$ and $(n,l)$,
\begin{eqnarray}
    \mathcal{R}_{\theta} =&& \begin{pmatrix}
        1 & 0 & 0 & 0 \\
        0 & c_\theta & -is_\theta & 0 \\
        0 & -is_\theta & c_\theta & 0 \\
        0 & 0 & 0 & 1 
\end{pmatrix},
\end{eqnarray}
where $c_{\theta} = \cos(\theta/2)$, $s_{\theta} = \sin(\theta/2)$. We also define the gate $W^{\prime\prime}=W^\prime\mathcal{R}^\dagger_{\theta}$.

Then, we can represent the unitary for our modified Dirac QCA as a circuit consisting of local unitaries as shown in Figure \ref{fig:ModCircuit}. The equivalence can easily be checked by using the transformations defined in Eq.~\eqref{eq:JW}. Again, one needs to restrict to a finite number of lattice sites with periodic boundary conditions for simulation on a quantum computer.

\section{Proof of the Theorem \ref{thm:theta}}\label{app:theorem}
Using $\Exp{i\hat{a}\cdot\vec{\sigma}\theta}= \mathbb{I}\cos(\theta)+i\hat{a}\cdot\vec{\sigma}\sin(\theta)$ where $\hat{a}$ is any unit vector, we can write $U_\text{mod}(p)$ from Eq.~\eqref{eq:modwalk} as
\begin{align}
     U_{\text{mod}}^{\text{QW}}(p) =\, &\big(\cos(m c^2\delta t)-i\sigma_x\sin(m c^2\delta t) \big) \nonumber\\
     &\times\big(\cos(p\,\delta x)-i\sigma_{-\theta}\sin(p\,\delta x)\big) \nonumber\\
     &\times\big(\cos(p\,\delta x)-i\sigma_{\theta}\sin(p\,\delta x)\big) .
     \label{eq:pauliexpand}
\end{align}

By expanding Eq.~\eqref{eq:pauliexpand}, we can rewrite $U_{\text{mod}}^{\text{QW}}$ in the form $r\mathbb{I}+iA$ where $A$ is a Hermitian operator and $r$ is a real number. Then, the real part of the eigenvalues of $U_\text{mod}(p)$ is simply $r$, which is given by
\begin{align}
 r &= \cos(mc^2\delta t)\cos^2(p\,\delta x)
 \nonumber\\
 &\qquad-\sin^2(p\,\delta x)\cos(mc^2\delta t\!-\!2\theta).
\end{align}
Then, defining $\phi \equiv \pi/2-\theta$, we can rewrite this as
\begin{align}
r&=\cos(mc^2\delta t)\cos^2(p\,\delta x)
\nonumber\\
 &\qquad
+\sin^2(p\,\delta x)\cos(mc^2\delta t\!+\!2\phi).
\end{align}

Writing the eigenvalues as $e^{- i\widetilde{E}_p \delta t}$ and considering the real part, we get
\begin{align}
    \cos(\widetilde{E}_p\delta t) &= \cos(mc^2\delta t)\cos^2(p\,\delta x)\nonumber\\
    &\quad+\sin^2(p\,\delta x)\cos(mc^2\delta t\!+\!2\phi). 
    \label{eq:inequality}
\end{align}

Now, we can choose $\theta$ such that $-\pi/2<mc^2\delta t\!+\!2\phi<\pi/2$. Moreover, $0\leq mc^2\delta t < \pi/2$. Hence, $\cos(mc^2\delta t\!+\!2\phi)>0$ and $\cos(mc^2\delta t)>0$. Then, it follows from Eq.~\eqref{eq:inequality} that $\cos(\widetilde{E}_p\delta t)> 0$. As $\widetilde{E}_p$ is restricted to $[-\pi/\delta t,\pi/\delta t)$, the theorem follows.

\section{The 3+1-D Case}\label{app:3d}
\subsection{The Dirac QW in 3+1-D}

Consider a quantum particle on a 3-D lattice with a four-dimensional internal degree of freedom. We suppose that the lattice is cubic with the lattice spacing being $\delta x$. The state of the particle is in the Hilbert space $\Hc=\Hc_{\Zb^3}\otimes\Hc_{\text{int}}$ where $\Hc_{\Zb^3} = \text{span}\{\ket{\vec{n}}|\vec{n}\in \Zb^3\}$ and $\Hc_{\text{int}}=\Hc_{\text{S}}^{\otimes2}\cong\mathbb{C}^4$. 
%Again, $\Hc_{\text{S}}$ is spanned by the orthonormal states $\ket{r}$ and $\ket{l}$.

Similar to the 1+1-D case, we can define the momentum eigenstates and the momentum operators as 
\be
\ket{\vec{p}} \!=\! \left(\frac{\delta x}{2\pi}\right)^{\tfrac{3}{2}}\sum_{\vec{n}} e^{i \vec{p}\cdot\vec{n}\delta x}\ket{\vec{n}} \;\text{and}\;
\vec{P} \!=\!\! \int_{\Theta} \!\!d^3 p \,\vec{p} \ketbra{\vec{p}},
\ee
where
\be
\Theta=[-\pi/\delta x,\pi/\delta x)^3.
\ee

For $j\in\{x,y,z\}$, we define the shift operators as
\be
    K_j = \Exp{-iP_j\sigma_j\delta x}
\ee
which act on $\Hc_{\Zb^3}\otimes\Hc_{\text{S}}$.
Then, we define the following operator acting on $\Hc_{\Zb^3}\otimes\Hc_{\text{int}}$,
\be
    T_j = \begin{pmatrix}
        K_j & 0 \\
        0 & K^\dagger_j
    \end{pmatrix},
\ee
with $j\in\{x,y,z\}$. 

We define $V$ and rewrite $T_j$ as
\be
   V = \Exp{-imc^2\beta\,\delta t} \;\text{and}\;  T_j = \Exp{-iP_j\alpha_j\delta x}  
\ee
where 
\be
    \alpha_j = \begin{pmatrix}
        \sigma_j& 0 \\
        0 & -\sigma_j
    \end{pmatrix}
     \;\text{and}\;
     \beta = \begin{pmatrix}
        0 & \mathbb{I} \\
        \mathbb{I} & 0
    \end{pmatrix}.
\ee

Then, the evolution of the particle in a single time-step $\delta t$, is given by the unitary 
\be
    U_{\text{3D}}^{\text{QW}} = VT_zT_yT_x.
\ee
This is the 3+1-D Dirac walk from \cite{farrelly2014causal}.
Consider the state $\ket{\psi_{\vec{p}}}=\ket{\vec{p}}\ket{s}$ where $\ket{s}$ is any state in $\Hc_{\text{S}}$. Then, consider
\be
U_{\text{3D}}^{\text{QW}}\ket{\psi_{\vec{p}}} = U_{\text{3D}}^{\text{QW}}\ket{{\vec{p}}}\ket{s} = \ket{\vec{p}}U(\vec{p})\ket{s},
\label{eq:evolnormalboundary3d}
\ee
where
\begin{align}
U(\vec{p})=&\Exp{-imc^2\beta\,\delta t}\Exp{-iP_z\alpha_z\delta x}\nonumber\\
&\quad\times\Exp{-iP_y\alpha_y\delta x}\Exp{-iP_x\alpha_x\delta x}.
\end{align}

\begin{figure*}
	\centering
	\begin{subfigure}{0.48\linewidth}
        \centering
		\includegraphics[width=\linewidth]{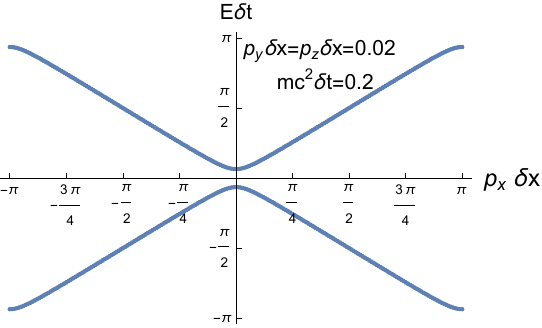}
		\caption{}
		\label{fig:3dLow}
	\end{subfigure}
 \hfill
	\begin{subfigure}{0.48\linewidth}
        \centering
		\includegraphics[width=\linewidth]{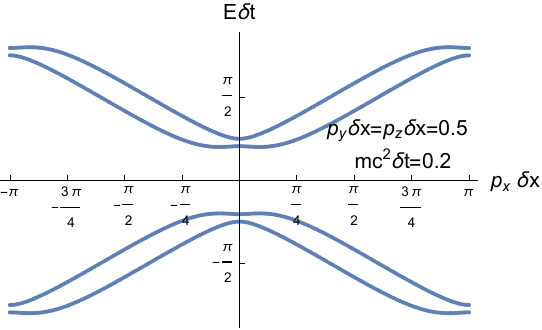}
		\caption{}
		\label{fig:3dHigh}
	\end{subfigure}
 \caption{Plots of the dispersion relation for the 3+1-D quantum walk.  Now, instead of two energy eigenvalues for each momentum, we have four energy eigenvalues. In Figure (a), while it seems that there are only two eigenvalues, we do have four distinct eigenvalues. The four eigenvalues approach two degenerate eigenvalues as any two components of momentum approach zero.}
 \label{fig:3DOGGraph}
\end{figure*}

Writing the eigenvalues of $U(\vec{p})$ as $\Exp{-i \tilde{E}_{\vec{p}} \delta t}$, we define the energy of the particle as $\tilde{E}_{\vec{p}}$. Again, defined this way, energy has a modularity of $2\pi/\delta t$. Restricting the energy from $-\pi/\delta t$ to $+\pi/\delta t$, we obtain a notion of positive and negative energies. Then, we again have two boundaries between the negative and the positive energy states ($E=0$ and $E=\pm\pi/\delta t$). The dispersion relation for fixed $m$, $p_y$ and $p_z$ has been plotted in Figure \ref{fig:3DOGGraph}. Here, we can see that in general we have four energy eigenvalues for each momentum eigenstate. However, in the appropriate limits where we would expect the quantum walk to behave like a free Dirac particle, the four values converge to just two values. 

Particularly, supposing $mc^2\delta t \ll 1$ and $p_j\delta x \ll 1$, we get
\begin{align}
    \!\!U_{\text{3D}}^{\text{QW}}\!\ket{\psi_{\vec{p}}} &=\! \ket{{\vec{p}}} U(\vec{p}) \ket{s}\nonumber \\ 
    &\approx\! \ket{{\vec{p}}}\Exp{\!-i(mc^2\beta\,\delta t+ \vec{p}\cdot\vec{\alpha}\,\delta x)}\!\ket{s}
    \nonumber \\
    &\approx\! \Exp{\!-i(mc^2\beta\,\delta t+ \vec{P}\cdot\vec{\alpha}\,\delta x)}\!\ket{\psi_{\vec{p}}}
    \label{eq:diraclowexpansion2d}.
\end{align}
Now, we define the effective Hamiltonian as $U_{\text{3D}}^{\text{QW}} = e^{-i H^{\text{eff}}\delta t}$. Comparing with (\ref{eq:diraclowexpansion2d}), we get the effective Hamiltonian for low momentum
\be
H^{\text{eff}} \approx mc^2\beta+\vec{P}\cdot\vec{\alpha}\frac{\delta x}{\delta t}.
\ee
where we have implicitly chosen a branch cut. Now, we fix $
   \delta t = \delta x/c$
for this QW to get
\be
    H^{\text{eff}} \approx mc^2\beta+c\vec{P}\cdot\vec{\alpha},
    \label{eq:effHamil3d}
\ee
which is the 3+1-D Dirac Hamiltonian. The energy eigenvalues of this Hamiltonian are $\pm E_{\vec{p}}$ where $E_{\vec{p}}=\sqrt{\abs{\vec{p}}^2c^2+m^2c^4}$.

While it is well-established that this walk behaves like Dirac particles in the low momentum limit, we shall now show that this is also the case around some other points in momentum space. Suppose $p_j\delta x \ll 1$ and $mc^2\delta t\ll1$.

Consider the state with momentum $\vec{q} = (p_x,\pm\pi/\delta x+p_y,\pm\pi/\delta x+p_z)^\mathbf{T}$. It is easy to see that 
\be
    U_{\text{Dirac}}^{\text{QW}}\!\ket{\psi_{\vec{q}}} =\! \ket{{\vec{q}}} U(\vec{q}) \ket{s} = \! \ket{{\vec{q}}} U(\vec{p}) \ket{s}.
\ee
Hence, we expect the behaviour to be of a Dirac particle of momentum $\vec{p}$. We find the energy eigenvalues to be approximately $\pm E_{\vec{p}}$. This is true if any two of the components of momentum are near $\pm\pi/\delta x$ and the other one is near $0$. Hence, we have some sort of fermion doubling.

Now, consider the case when any two components are near $0$ and the other one is near $\pm\pi/\delta x$ or all three components are near $\pm\pi/\delta x$. For particles with momentum either $\vec{q} = (p_x,p_y,\pm\pi/\delta x+p_z)^\mathbf{T}$ or $\vec{q} = (\pm\pi/\delta x+p_x,\pm\pi/\delta x+p_y,\pm\pi/\delta x+p_z)^\mathbf{T}$, we get 

\be
    U_{\text{Dirac}}^{\text{QW}}\!\ket{\psi_{\vec{q}}} =\! \ket{{\vec{q}}} U(\vec{q}) \ket{s} = \! (-1)\ket{{\vec{q}}} U(\vec{p}) \ket{s}.
\ee

Thus, around these points, the quantum walk again behave like a free Dirac particle but with a global phase factor and energies approximately $-\pi/\delta t + E_{\vec{p}}$ and $\pi/\delta t - E_{\vec{p}}$. 

\begin{figure*}
	\centering
	\begin{subfigure}{0.48\linewidth}
        \centering
		\includegraphics[width=\linewidth]{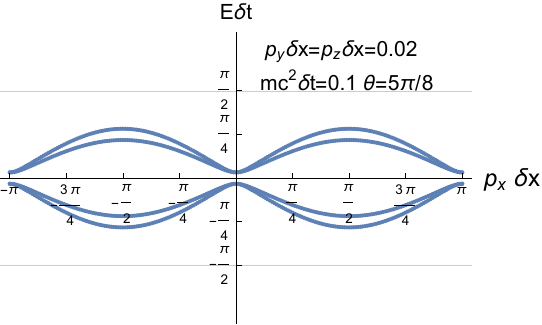}
		\caption{}
		\label{fig:fig:3DModGraph}
	\end{subfigure}
 \hfill
	\begin{subfigure}{0.48\linewidth}
        \centering
		\includegraphics[width=\linewidth]{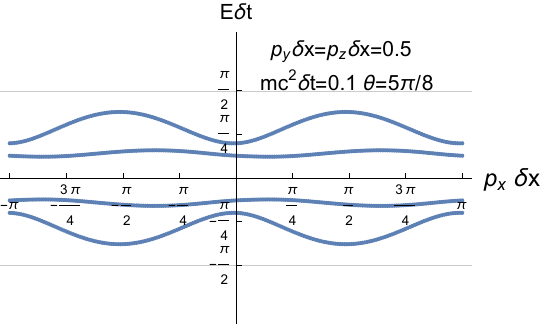}
		\caption{}
		\label{fig:3DModGraph2}
	\end{subfigure}
 \caption{Plots of the dispersion relation for the 3+1-D modified quantum walk. We can see that the energy eigenstates are pushed far enough from the problematic boundary such that the absolute energy is never greater than $\pi/(2\,\delta t)$.}
 \label{fig:3DModGraph}
\end{figure*}

If we look at the massless case ($m=0$), we find another additional point in momentum space with interesting behaviour. Consider a particle with momentum $\vec{p}$ such that $p_j \delta x \ll 1$ again. Then, the quantum walk simulates a right-moving and a left-moving Weyl particle with energies approximately $\pm\abs{\,\vec{p}\,}$. Apart from the same points as discussed for the massive case, there is another point in the momentum space around which the quantum walk behaves like two free Weyl particles with opposite chiralities. This is the case if all of the components of the momentum vector are of the form $\pm\pi/(2\,\delta x) + p_j$  where $p_j\delta x \ll 1$. Then, the energy eigenvalues are approximately $\pm\abs{\,\vec{p}\,}c$, $-\pi/\delta t+\abs{\,\vec{p}\,}c$ and $\pi/\delta t-\abs{\,\vec{p}\,}c$. Thus particles near these points are present near the boundaries between the negative and positive energy particles. Note that, for the massive case ($m>0$), while we do not expect the quantum walk around these point to simulate a free Dirac particle, we do expect  a particle near these points to be near one of the two boundaries. However, the exact behaviour around this point for the massive case is yet to be studied.

Finally, one can upgrade this model to a QCA. Again, there are high energy particles that behave like Dirac particles and live near the second boundary between the positive and negative energy states ($E=\pm\pi/\delta t$). Hence, similar to the 1+1-D case, we would expect pair creation at the second boundary if we were to fill up all the `negative' energy states. Again, one way to solve this problem is to modify the walk such that the available energy eigenstates are pulled away from the problematic boundary.

\subsection{Modified 3+1-D Dirac QW}

Now, consider the following unitaries
\begin{align}
    \widetilde{K}_{z,\theta} = R_{x,\theta} K_z R_{x,\theta}^\dagger,  \\
    \widetilde{K}_{y,\theta} =  R_{z,\theta} K_y R_{z,\theta}^\dagger,  \\
    \widetilde{K}_{x,\theta} = R_{y,\theta} K_x R_{y,\theta}^\dagger.
\end{align}
Consider 
\be
 \widetilde{T}_{j,\theta} = \begin{pmatrix}
         \widetilde{K}_{j,\theta} & 0 \\
        0 & \widetilde{K}_{j,\theta}^\dagger
    \end{pmatrix}
\ee
and, as before, 
\be
    V = \Exp{-imc^2\beta\,\delta t}.
\ee
Then, we introduce the quantum walk given by
\begin{align}
    U_{\text{3D mod}}^{\text{QW}} =& V \widetilde{T}_{z,-\theta} \widetilde{T}_{z,\theta}
    \nonumber\\ &\times \widetilde{T}_{y,-\theta} \widetilde{T}_{y,\theta}\widetilde{T}_{x,-\theta} \widetilde{T}_{x,\theta}.
\end{align}
 
Setting $\delta t = 2\cos(\theta)\delta x/c$, we can check that this quantum walk simulates a Dirac particle for $p_j\delta x\ll1$ and $mc^2\delta t\ll 1$. Defining energy in a similar way as before, we have plotted the dispersion relation in Figure \ref{fig:3DModGraph}. Similarly to the one-dimensional case, we expect that for any $0\leq mc^2\delta t < \pi/2$, we can choose an appropriate value of $\theta$ such that the gap between the positive and negative energy states is larger than $\pi/\delta t$. 

%\printbibliography
\bibliographystyle{quantum}
\bibliography{myref}
\end{document}